\documentclass[doc,a4paper,floatsintext]{apa7}
\usepackage[utf8]{inputenc}
\usepackage[T1]{fontenc}
\usepackage[style=apa, backend=biber, url=true]{biblatex}
\usepackage{booktabs}
\usepackage{graphicx}
\usepackage[export]{adjustbox}
\usepackage{listings,multicol}
\usepackage{varioref}
\usepackage{hyperref}
\usepackage{cleveref} 
\usepackage{placeins}
\usepackage{tablefootnote}
\usepackage{xcolor}
\usepackage{array}
\usepackage{calc} 
\usepackage{etoolbox}
\usepackage{bookmark}
\IfFileExists{xurl.sty}{\usepackage{xurl}}{} 
\hypersetup{
  colorlinks=true,
  linkcolor={blue},
  filecolor={Maroon},
  citecolor={blue},
  urlcolor={blue},
  pdfcreator={Christophe Pallier <christophe@pallier.org>}}

\title{Goxpyriment: A Go Framework for Behavioral and Cognitive Experiments}
\shorttitle{Let's go experiment!}

\authorsnames{Christophe Pallier, Julie Bonnaire, Marie-France Fourcade}
\authorsaffiliations{Cognitive Neuroimaging Unit, CNRS INSERM CEA}

\authornote{\addORCIDlink{Christophe Pallier}{0000-0002-3324-666X} Correspondence should be addressed to  C. Pallier, Cognitive Neuroimaging Unit, Neurospin, Gif-sur-Yvette, F-91191, France. E-mail: christophe@pallier.org. I would like to thank Luca Bonatti and Raphael Bordas for alpha testing this software. The authors have no conflicts of interest to disclose.}


\abstract{We introduce \texttt{goxpyriment}, an open-source software
  framework for programming behavioral and cognitive experiments in
  the Go programming language. It addresses two recurrent practical
  difficulties of existing Python-based tools: the runtime-environment
  complexity that complicates deployment across laboratories, and the
  unpredictable pauses that automatic memory management can introduce
  during timing-critical sections. Because Go is a compiled language
  that can natively embed assets (e.g., graphics, audio files, and
  stimulus lists), \texttt{goxpyriment} compiles an entire experiment
  into a single, self-contained executable with no runtime
  dependencies, so that distributing an experiment amounts to copying
  one file. Users can just click on the executable within their file
  browser and do not need to open a terminal and use the command line
  (but they can if they want/need to).  The programming interface is
  inspired by Expyriment \parencite{krause2014}. It provides an array
  of visual stimuli (text, shapes, images, Gabor patches, moving dot
  clouds, \ldots) and audio capabilities (WAV playback and tone
  generation). For precise timing, response times are obtained by
  using operating-system event timestamps recorded at
  hardware-interrupt time rather than by continuous polling, and the
  garbage collector can be suspended for the duration of a trial. TTL
  triggers for EEG, MEG and fMRI are supported on a range of devices,
  and eye trackers (EyeLink, Tobii Pro) are driven through a
  vendor-neutral interface. A characterization of the resulting timing
  behaviour with the hardware-based protocol of the timing mega-study
  \parencite{bridges2020} is in progress. The same source
  code also compiles to WebAssembly, so that an experiment can be run
  in a web browser from a URL when browser-grade timing is
  acceptable. The framework ships with over sixty example experiments
  and some thirty-five demonstrations that serve as reusable templates
  for common paradigms; all are available as pre-built programs for
  Windows, macOS and Linux, and most as browser
  versions. \texttt{goxpyriment} is released under the Apache License
  2.0 and is freely available at
  \url{https://github.com/chrplr/goxpyriment}.}

\begin{document}
\maketitle

\section{Introduction}\label{introduction}

Experiment-control software provides the foundation for empirical
behavioral and cognitive neuroscience: the temporal precision of
stimulus delivery and response recording, the ergonomics of experiment
programming, and the ease of deploying experiments to laboratory
hardware each affect the quality and reproducibility of empirical
data. Over the past two decades, Python has become the dominant
language for experiment programming, primarily through libraries such
as PyEPL \parencite{pyepl_2007}, VisionEgg \parencite{visionegg_2008},
Expyriment \parencite{krause2014} and PsychoPy
\parencite{peirce2019}. Before that, MATLAB with the Psychophysics
Toolbox \parencite{brainard1997, kleiner2007} was the de facto
standard in many laboratories and is still in use in many of
them. More recently, web-based frameworks such as jsPsych
\parencite{deleeuw2015, deleeuw2023} and lab.js
\parencite{henninger2022} have enabled large-scale online data
collection.

Each class of tool involves tradeoffs. MATLAB, extended with the
Psychtoolbox, is convenient to generate visual or auditory stimuli and
offers exellent timing on well-configured hardware. Yet Psychtoolbox
scripts written on Windows most often do not work out of the box on
Linux (and vice-versa). Also, MATLAB can be costly for institutions
that cannot obtain academic licenses (as is the case for some
non-degree-granting research institutions).  Python tools provide a
more open programming environment with a large scientific ecosystem.
Python-based experiments can also achieve high-precision timing, for
example by relying on the psychtoolbox Python module which provides
pieces of the Psychtoolbox-3 (see
\url{https://pypi.org/project/psychtoolbox/}). However, the use of
Python introduces practical difficulties. For instance, deploying an
experiment to a computer that does not already have a compatible
Python environment requires package management (pip, venv, conda, uv,
\ldots) and can be a source of breakage across operating-system
versions. For example, installing the \texttt{psychopy} module under
Linux fails while compiling the wxPython package when no precompiled
wheel is available. Second, Python's garbage collector can introduce
unpredictable pause times that corrupt the timing of
stimulus presentation and response recording during the critical trial
loop. Web-based tools have an obvious advantage in terms of facility
of deployment, but browser-mediated stimulus presentation is subject
to event-loop latency that makes precise timing difficult
\parencite{bridges2020, anwylirvine2021}.

The Go programming language \parencite{golang} offers a different set
of tradeoffs. Go is a deliberately small language with a limited
feature set and few alternative ways to express a given operation,
which tends to make programs easy to read and to reason about
\parencite{donovan2016}. It is statically typed, and its conventions
encourage explicit error checking, so that many mistakes surface at
compile time rather than at run time, which is a huge time saver for
the programmer.  Moreover, two properties of Go make it an
attractive language to program and distribute experiments.  First, Go
(cross-)compiles to native machine code and can produce self-contained
binaries that embed the stimuli and experiment lists and have thus no
runtime dependencies. Second, its garbage collector is concurrent and
can be suspended for the duration of a timing-critical section. These
properties motivated our choice of Go as the basis for a behavioral
experiment library that prioritizes deployment simplicity and timing
control.

The present article describes the design and capabilities of
\texttt{goxpyriment}, and details its timing architecture. We highlight four
contributions: (1) zero-dependency deployment via single compiled
binaries that simplify distribution across laboratory hardware; (2) a
timing architecture that obtains response times from operating-system
event timestamps and that can suspend garbage collection during
trials, whose behaviour is being measured with the protocol of the timing
mega-study \parencite{bridges2020}; (3) a single code base that runs
natively on Linux, macOS and Windows and, unchanged, in a web browser
through WebAssembly; and (4) a gallery of more than sixty
experiments and thirty-five demonstrations, each with its reference
and a pre-built binary.

{\small
\begin{lstlisting}[float=tp, label=paritycode, caption={Example of a Parity Decision experiment}, lineskip=-5pt, breaklines=true, breakatwhitespace=true, basicstyle=\small\ttfamily]
package main

import (
	"strconv"  // to convert int -> string

	. "github.com/chrplr/goxpyriment/control"
	"github.com/chrplr/goxpyriment/design"
	"github.com/chrplr/goxpyriment/stimuli"
)

func main() {
	exp := NewExperimentFromFlags("Parity Decision",
		Black, White, 32)
	defer exp.End()

	Instructions := "When you'll see a number, your task to decide, " +
		"as quickly as possible, whether it is even or odd.\n\n" +
		"if it is even, press 'F'\n\nif it is odd, press 'J'"

	Targets := []int{0, 1, 2, 3, 4, 5, 6, 7, 8, 9}
	NTrialsPerTarget := 2
	trials := design.RepeatList(Targets, NTrialsPerTarget)
	design.ShuffleList(trials)

	EvenResponse := K_F
	OddResponse := K_J
	respKeys := []Keycode{EvenResponse, OddResponse}

	cross := stimuli.NewFixCross(25, 2, DefaultTextColor)

	// creates a map number -> image
	Image := make(map[int]*stimuli.TextLine)
	for _, num := range Targets {
		Image[num] = stimuli.NewTextLine(strconv.Itoa(num),
			0, 0, DefaultTextColor)
	}

	exp.AddDataVariableNames([]string{"number", "key", "rt",
		"correct"})

	exp.Run(func() error {
		exp.ShowInstructions(Instructions)

		for _, num := range trials {
			exp.Blank(1500)
			exp.ShowTimed(cross, 500)
			key, rt, err := exp.ShowAndGetRT(Image[num], respKeys, -1)
			if err != nil {
				return err // ESC: end the run, without recording this trial
			}

			correct := (num%2 == 0) == (key == EvenResponse)
			exp.Data.Add(num, key, rt, correct)
			if !correct {
				exp.Audio.PlayBuzzer()
			}
		}
		return EndLoop
	})

	exp.ShowEndMessage("Experiment complete. Thank you!\n\n" +
		"Press any key to exit.")
}
\end{lstlisting}
\par}

\section{Framework Design}\label{framework-design}

\subsection{Foundation: SDL3}

\texttt{Goxpyriment} relies on the Simple DirectMedia Layer, a
cross-platform development library designed to provide low-level
access to audio, keyboard, mouse, joystick, and graphics hardware
\parencite{sdl3}.  SDL has been used to develop thousands of
commercial games as well as many emulators. Bindings for many
programming languages exist, notably for Python with the \verb|pygame|
library \parencite{pygame}. Because of its large user and developer
bases, SDL is actively maintained and keeps up with modern hardware.
Goxpyriment uses the latest version of this library, SDL3, through the
go-sdl3 bindings \parencite{gosdl3}.

\subsection{Single-Binary Deployment (Zero Dependencies)}\label{single-binary-deployment}

Go compiles programs into a single executable with no external runtime
dependencies. Crucially, the \texttt{go-sdl3}
package embeds the pre-built SDL3 runtime libraries \parencite{sdl3}
directly into the executable, so that no library needs to be installed
separately on the target machine.  In addition, the stimuli,
experimental lists, fonts, and other assets can be embedded in the
binary too. As a result, distributing an experiment requires copying
just one file; there is no need to manage virtual environments,
package versions, or interpreter installations across laboratory
computers. This substantially reduces the operational overhead of
installing experiments on multiple machines or sharing materials with
collaborators.

Another important consequence of the executable carrying its own copy of
the SDL3 libraries (embedded in the binary and loaded from it at run time) is that the user does not have to worry about
potential changes in future versions of the library which could impact
the behavior of their program. This promotes reproducibility.

Because Go cross-compiles without a C toolchain, every example in the
repository is built for Windows (x86-64), macOS (Apple silicon) and
Linux (x86-64 and arm64, including the Raspberry~Pi) from a single
Linux machine by the continuous-integration pipeline, and published
both as one archive per platform and as one small archive per
experiment (\url{https://downloads.pallier.org/builds/latest/}). The
Windows builds are linked as GUI programs so that no console window
opens beside the experiment. The binaries are not code-signed, so
macOS Gatekeeper and Windows SmartScreen show a warning on first
launch; the documentation gives the one-line workaround for each.

\subsection{The same experiment in a web browser}\label{browser}

The same source code compiles, unchanged, to WebAssembly
(\texttt{GOOS=js GOARCH=wasm}). SDL3 and its font, image and audio
companions are compiled by Emscripten once and shipped as a pre-built
blob in a fork of the \texttt{go-sdl3} bindings, so an experimenter
needs no Emscripten installation: \texttt{make wasm-NAME} produces a
directory that any static web server can host. In the browser, the
whole \texttt{Experiment} API is available: keyboard and mouse
responses carry event timestamps, audio plays through Web Audio, and
\texttt{Screen.Update()} parks on the browser's
\texttt{requestAnimationFrame}, which is the closest thing a browser
has to VSYNC. Session parameters that the desktop version takes from
the command line or from the participant-information dialog are read
from the URL query string (\texttt{?s=3}), and at the end of the
session the participant's browser downloads a single \texttt{.zip}
holding the data CSV and the \texttt{-info.txt} companion, byte for
byte what a native run writes.

For online data collection, the browser build can also be deployed on a
JATOS server \parencite{lange2015}, which hosts the study, distributes
links to participants and stores their results. \texttt{make jatos-NAME}
packages an experiment, again without changes to its source, as a
JATOS study archive (\texttt{.jzip}) that the experimenter imports
through the JATOS web interface. Parameters appended to the study link
reach the experiment as they would on the command line, and each call
to \texttt{exp.Data.Save()} sends the rows recorded so far to the
server, so that a participant who closes the tab halfway through
leaves behind every block completed until then. The researcher
retrieves the same CSV and \texttt{-info.txt} files as from a lab
session with JATOS's \emph{Export Results}.

Two honest caveats. First, a browser cannot offer the display-stack
guarantees of the native build: there is no hardware retrace clock,
no real-time priority, and we have not validated pixel onset with a
photodiode in a browser. What we did measure is the timestamp
resolution and the frame loop: SDL and input-event timestamps tick at
about 5~$\mu$s when the page is served with cross-origin-isolation
headers (about 100~$\mu$s otherwise), and a 299-frame presentation
loop in headless Chrome ran at 16.666~ms (SD 0.12~ms) with no dropped
frame. Paradigms that hinge on sub-millisecond onsets (rapid RSVP,
subliminal priming) should therefore run natively; choice
reaction-time, memory and attention tasks are well served. Second,
some capabilities have no browser counterpart: hardware triggers,
gamepads, audio recording, video playback and multi-display
fullscreen, and all assets must be embedded (the browser has no file
system). Ninety-two of the hundred and four examples satisfy these
constraints and are published as browser versions; each example's
README links to its own.

\subsection{Programming interface: a simple example}\label{example}

Listing~\vref{paritycode} provides an example of a simple experiment
implementing a parity decision task where participants have to decide
if a number is even or odd.

\begin{itemize}
\item Line 12: An \texttt{Experiment} object that manages the
window, renderer, and audio subsystems must be created
early. Calling it also parses the command-line to get the subject's
code and whether the experiment should run in fullscreen mode or not.

\item Lines 16--36: This part of the code prepares the stimuli. Lines
31--36 are the most Go-specific: they create a map (a.k.a. associative
array or dictionary in other programming languages) associating each
number to a graphic texture. By pre-building these textures, we ensure
that they will be displayed as quickly as possible in the presentation
loop (we could have called \texttt{stimuli.NewTextLine()} in the loop,
saving a few lines at the cost of recomputing the texture at each
iteration).

\item Lines 41--59: This part of the code deals with the presentation
  loop. It is wrapped in a \texttt{Run} call that catches quitting
  events: if the participant presses the Escape key or closes the
  window, the library catches the resulting internal signal, safely
  flushes the data file to disk, and shuts down the SDL subsystems,
  ensuring that data is preserved even if a session is aborted
  prematurely. Line 47, the function \texttt{ShowAndGetRT()} presents
  the visual stimulus, waits for one of the given keys, then returns
  the reaction-time; lines 48--50 return its error when the participant
  quits, so that the interrupted trial is not recorded.
\end{itemize}

\subsection{Timestamps, Garbage Collection Control, ...}

Response timing in \texttt{goxpyriment} exploits SDL3 OS-level event
timestamps. The \verb|ShowAndGetRT| function used in the parity example is
a wrapper for:

\noindent%
\begin{lstlisting}[numbers=none,basicstyle=\small]
  exp.Keyboard.Clear()  // discard presses made before the onset
  onset,_ := exp.ShowTS(stim)
  key,eventTS,_ := exp.Keyboard.GetKeyEventTS(keys,
                                              timeoutMs)
  rt := int64(eventTS - onset) / 1_000_000
\end{lstlisting}

\texttt{ShowTS()} displays the stimulus \texttt{stim}, blocking until
the next screen flip when VSYNC is enabled, and returning a
time-stamp for that event (in ns).  When VSYNC is disabled,
\texttt{ShowTS()} returns immediately, supporting variable
refresh rate (VRR) monitors.

\verb|GetKeyEventTS()| scans the keyboard and returns the code of the
key pressed and a timestamp (in ns) for it. This timestamp is
populated by the OS input subsystem at hardware-interrupt time
(nanoseconds since SDL initialization).  Because SDL events carry
their timestamp (information that pygame discards), the event remains in
the queue with its original timestamp regardless of how much
computation occurs between stimulus onset and response collection.  By
comparing the response timestamp with the \texttt{ShowTS()} timestamp,
reaction times can be computed without interpreter overhead
(hardware-level lags at the keyboard or mouse can, of course, still be
present). An advantage of this approach is that there is no need to
poll the HID device continuously to detect events: a complex sequence
of stimuli can be programmed and every event remains available, with
its original timestamp, to be processed later when needed.

To reduce the likelihood of latencies, goxpyriment provides additional functions:

\begin{description}
\item[Explicit Garbage Collection Control] Go, like Python or
  JavaScript, has a memory garbage collector. This can introduce
  unpredictable pauses that corrupt timing. However, unlike Python or
  Javascript, Go allows the programmer to suspend the garbage
  collector.  This can be used in \texttt{goxpyriment}'s
  timing-critical loops to prevent garbage-collection during a trial.

\item[High-Precision Streams] For paradigms with dense, high-speed
  stimulus sequences (e.g., retinotopic stimulation),
  \texttt{goxpyriment} provides specialized stream functions such as
  \texttt{PresentStreamOfImages}. These functions disable garbage
  collection, drain the event queue, and busy-wait at the sub-frame
  level to ensure that each stimulus onset is scheduled accurately. A
  \texttt{TimingLog} records the target and actual onsets for post-hoc
  verification, as well as the timing of keys pressed during the
  presentation of the stream.

\item[Real-Time Scheduling Priority] On Linux, an experiment built with
  \texttt{goxpyriment} requests the \texttt{SCHED\_FIFO} real-time
  scheduling policy for itself at start-up (priority 50 by default),
  rather than relying on the user to remember a \texttt{chrt} prefix on
  the command line --- which is in any case impossible when the program
  is launched by clicking its icon. The request is made with
  \texttt{sched\_setscheduler()} on the main thread, to which the Go
  runtime has been pinned, so that the elevation applies to the thread
  that runs the trial loop. Under the default (\texttt{SCHED\_OTHER})
  policy the kernel may deschedule the process at any point, including
  between a stimulus onset and the trigger that marks it; this is the
  dominant source of the rare, large latencies that otherwise survive
  the measures above.

  The request is not silently mandatory. It requires a system-level
  grant (an \texttt{rtprio} limit for the user, set in
  \texttt{/etc/security/limits.d/}), and if that grant is absent the
  program reports the reason and continues at normal priority rather
  than refusing to run. It can be disabled with the
  \texttt{-no-realtime} flag --- useful when debugging, since a
  breakpoint hit on a real-time thread can leave the desktop
  unresponsive --- and the priority can be changed with
  \texttt{-realtime-priority}. The policy actually obtained is recorded
  in the session's system report, so that a data set carries evidence of
  the conditions under which it was collected rather than of the
  conditions that were intended.
\end{description}

\subsection{Triggers}

To communicate with external recording equipment, such as fMRI, EEG,
and MEG systems, \verb|Goxpyriment| exposes every TTL device through
one eight-line input/output interface, so that an experiment written
against one device runs against another by changing a flag. The
current version drives the parallel port and GPIO lines (Linux), the
DLP-IO8 (\url{https://github.com/chrplr/dlp-io8-g}) and DLP-IO20
USB-serial boxes, the Adafruit FT232H (Linux; in pure Go, without the vendor
library), the LabJack T4 over Ethernet
(\url{https://labjack.com/products/labjack-t4}, without its SDK), the
NEUROSPEC MMBT-S, and an Arduino Mega~2560 running a custom firmware
that also timestamps button presses on the device
(\url{https://github.com/chrplr/neurospin-meg-ttl-box}). Beyond TTL,
the package talks to a generic serial port, to EGI NetStation over its
network protocol, and to a networked video recorder. Adding support
for new hardware is facilitated by the availability of the source code
for the current devices.

How the pulse is issued matters as much as which device issues it.
Go makes asynchronous execution trivial---prepending the \verb|go|
keyword to a call runs it concurrently and returns at once---and it is
tempting to fire every trigger that way. But a goroutine is scheduled
at the runtime's convenience: measured against the flip timestamp, a
rising edge dispatched through a goroutine at normal priority under
load arrived 0.73~ms late with about 1~ms of spread, while the same
edge raised synchronously on the statement after the flip, at
real-time priority, arrived at a median of 34~$\mu$s (maximum
37~$\mu$s). \texttt{goxpyriment} therefore provides
\texttt{FireTriggerSync}, which raises the line on the calling thread
and defers only the falling edge, which carries no information, to a
goroutine. The fire-and-forget form remains the right one for a block
marker or an end-of-run flag, where the edge is not the measurement.

The device itself costs little. Firing the same pulse train on two
devices from one schedule and reading the edges on an oscilloscope,
the DLP-IO8's edge lagged the FT232H's by 179.4~$\mu$s (SD 5.7, $n =
322$) on one machine and the parallel port's by 185.9~$\mu$s (SD 12.6,
$n = 900$, worst pulse 254.8~$\mu$s) on another: the USB-serial path
adds a constant of about 180~$\mu$s and a jitter an order of magnitude
below a millisecond. (This is the delay between two devices on one
schedule, not the delay of a device's edge relative to the flip
timestamp.)

\subsection{Eye tracking}\label{eyetracking}

Eye trackers are reached through a vendor-neutral \texttt{Tracker}
interface---open, calibrate, start and stop recording, mark, read the
latest sample or drain the buffered samples and the tracker's own
fixation, saccade and blink events---with implementations for the
SR~Research EyeLink and the Tobii~Pro range, a null tracker so that an
experiment runs unchanged without hardware, and a simulated tracker
driven by the mouse for development. Both vendors publish their
interface only as a C library or a native Python extension, and
linking either would cost \texttt{goxpyriment} its pure-Go build,
its cross-compilation and its browser target. The vendor SDK therefore
runs in a small Python \emph{bridge} process that the experiment talks
to over a loopback TCP socket, one JSON object per line. This is the
shape PsychoPy's ioHub arrived at for a different reason: there the
separate process keeps the interpreter lock out of the sampling loop;
here it keeps the C compiler out of the build. Supporting a further
tracker is a new bridge script, not a new Go package.

Calibration is one call, \texttt{exp.CalibrateTracker}, behind which
two procedures hide. The EyeLink draws its own calibration targets in
a window it owns, after which \texttt{goxpyriment} reclaims the
display and keyboard focus; the Tobii SDK draws nothing, so
\texttt{goxpyriment} draws the targets itself, with flip-timestamped
onsets, and drives the tracker's collect-and-compute protocol point
by point. The outcome of either is written to the session's
\texttt{-info.txt}.

The bridge is deliberately kept out of the timing path. Sending a
message through it and waiting for the tracker's acknowledgement took
a median of 600--719~$\mu$s and up to 1.2~ms over three runs on an
EyeLink at 2000~Hz---small against a frame, but not a way to time a
stimulus. Stimulus onsets are therefore marked by a TTL edge into the
tracker's digital input (with \texttt{FireTriggerSync}, as above),
and bridge messages serve only to label trials. The relation between
the tracker's clock and the experiment's can be measured NTP-style
(\texttt{Bridge.Sync} keeps the fastest of several round trips); called
before and after a run, as the bundled eye-tracker tests do, it gives the
offset with its round-trip uncertainty and the drift between the two clocks,
which those tests record in the data file's companion.

\subsection{Miscellaneous additional features}

This parity decision example only demonstrates a tiny part of the features
provided by the framework.

For example, it is possible to build the experiment around a
hierarchical structure of blocks and trials, over which the program
can iterate, following a pattern inspired by Expyriment
\parencite{krause2014}. Each trial is a collection of factors that can
be randomized and logged. Data collection is handled by an
\texttt{exp.Data} object, which writes \emph{two} files per session. The first
is the trial data as plain CSV, with no comment or metadata lines, so it opens
directly in a spreadsheet program or a statistics package without preprocessing.
The second is a companion \texttt{-info.txt} holding everything about the
conditions the session ran under: the machine, operating system, kernel and
desktop environment; every GPU and its driver; the sound server and its version;
the display mode, audio device and buffer size that were actually opened; the
scheduling policy obtained; and the participant details. Splitting them this way
means the analysable file needs no cleaning and the provenance is still
recorded --- and recorded automatically, rather than depending on someone
writing it down. An example appears in the
\hyperref[app:timing]{Supplementary Material} (Figure~\ref{fig:infofile}).

Beyond the stimuli visible in the example, the library provides
text-entry and menu widgets, thick poly-lines and filled polygons,
sprite sheets, random-dot stereograms, Gabor patches, dot clouds and
VSYNC-locked animation loops (drifting gratings, moving dot clouds),
and the \emph{stream} functions for RSVP and mixed audio-visual
sequences. Video plays in a frame-exact format (\texttt{.gv}) whose
onsets are locked to the retrace, and a \texttt{media} package
schedules several movies on one master clock. Naming latencies are
measured with a software voice key on the microphone input, which
returns the vocal onset timestamp and saves the recording per trial.
For psychophysics, a \texttt{staircase} package implements the
up-down \parencite{levitt1971} and QUEST \parencite{watson1983}
procedures, and a \texttt{units} package converts between pixels,
centimetres and degrees of visual angle for a described monitor and
viewing distance.

\begin{figure}[t]
\centering
\begin{tabular}{l}
  \textbf{A.}\\
  \includegraphics[width=0.5\textwidth,valign=t]{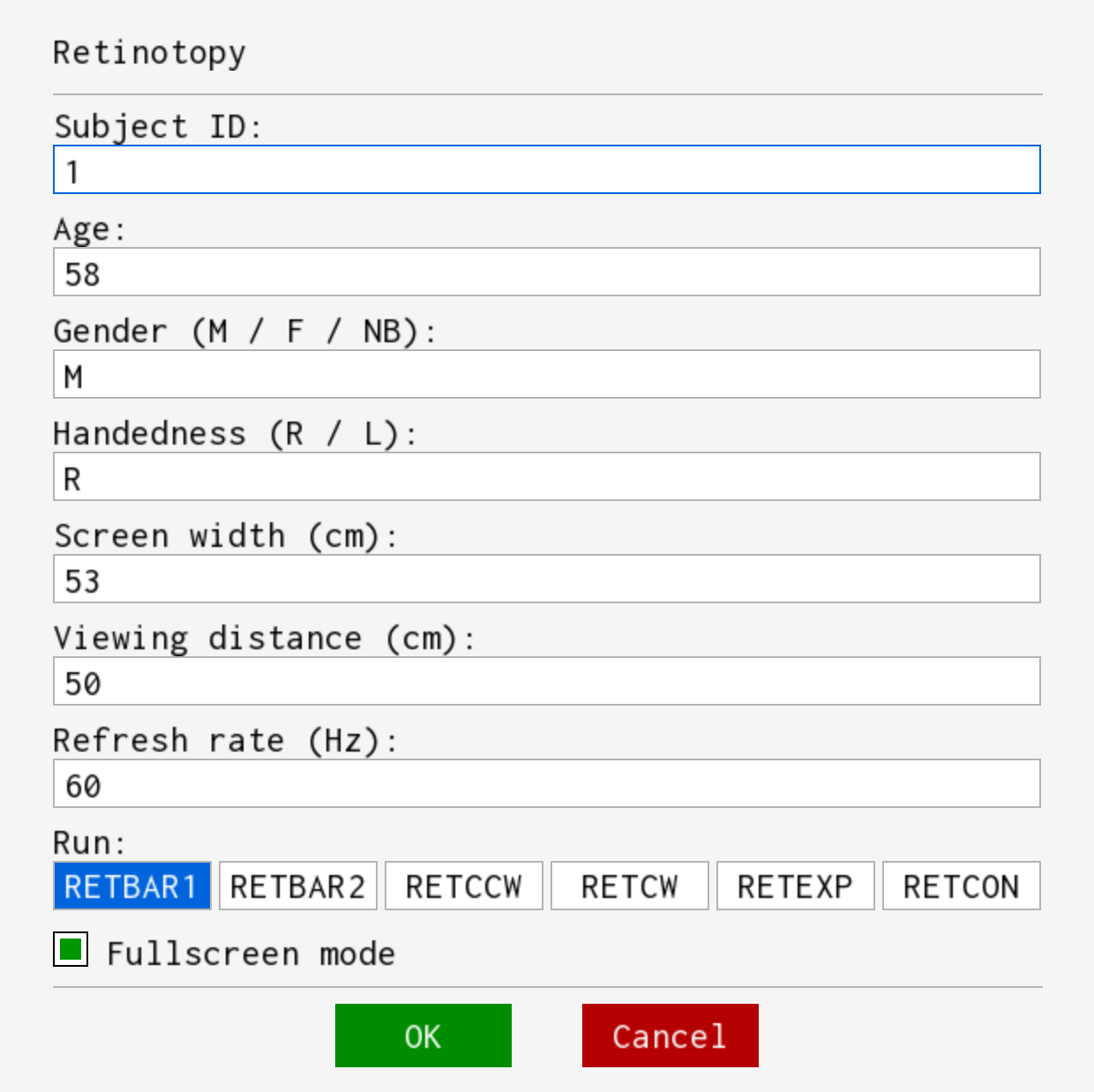} \\[.5cm]
  \textbf{B.}\\
  \includegraphics[width=0.5\textwidth,valign=t]{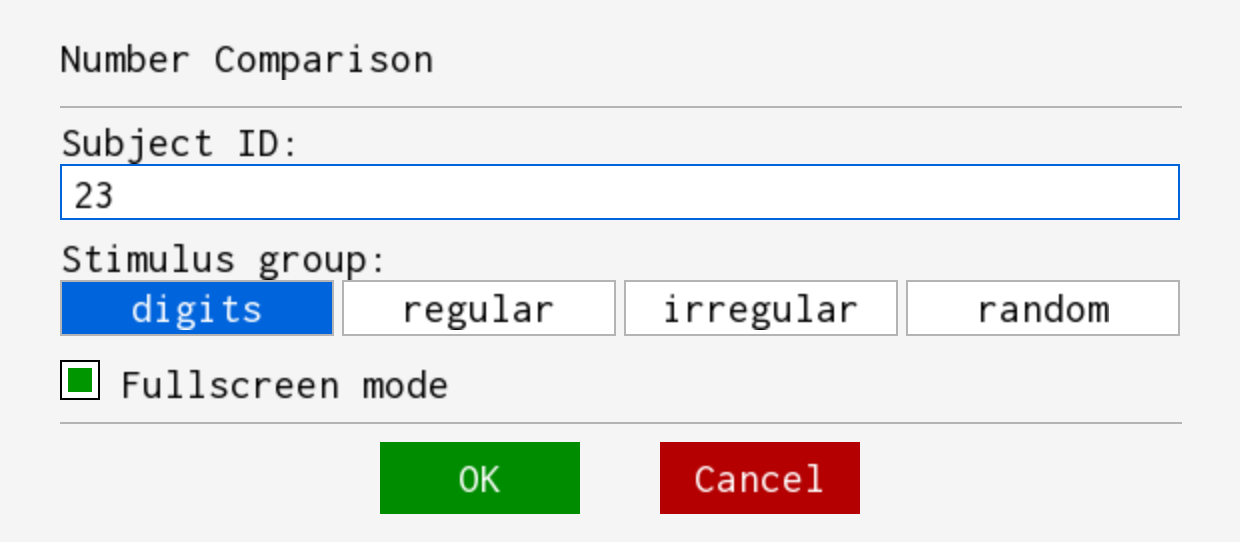} \\
\end{tabular}
\caption{Examples of \texttt{GetParticipantInfo} dialogs for
  \emph{Retinotopy} (A.) and \emph{Number-Comparison} (B.), two paradigms
  provided in the examples accompanying the library. These UIs makes it
  practical for the experimenter to choose among a small set of
  options without typing, which reduces setup errors when launching an
  experiment from a desktop shortcut or file manager. Note that which
  fields appear in the UI is configurable. It is also possible to
  pass the information as arguments on the command-line and skip these
  UIs.}
\label{fig:getinfo}
\end{figure}

Another thing which is not transparent from the listing of the parity
decision example is the fact that \texttt{goxpyriment} provides a
graphical \texttt{GetParticipantInfo} dialog that opens (optionally)
before the main experiment window (see Fig.~\ref{fig:getinfo}). This
allows the experimenter to input subject IDs, select experimental
conditions from discrete menus, and toggle settings like fullscreen
mode. These values are automatically recorded in the session's
\texttt{-info.txt} companion file.

\section{Timing Validation}\label{timing-verification}

This section used to contains a long an convoluted analysis of many timing tests.
I have decided to clean it and redo a campaign of measurments.

Work in progress...

\section{Gallery of Examples}

The repository carries, at the time of writing, 69 experiments, 35
demonstrations and 27 hardware tests, catalogued at
\url{https://chrplr.github.io/goxpyriment/GalleryOfExamples/}. An
\emph{experiment} implements a published paradigm and records
behavioural data; a \emph{demonstration} shows one feature (an
illusion, a widget, a stream) and measures nothing; a \emph{test} is
a program whose output is analysed to check the performance of the
machine it runs on (the timing suite, the trigger-device comparison, the
eye-tracker checks). Every directory declares its category, a
one-line description and its literature reference in a small metadata
file from which the gallery is generated, and carries a README.

The experiments span the classic repertoire of the field: attention
(Posner cueing and the ANT, attentional blink, change blindness,
multiple-object tracking, visual search, the psychological refractory
period); memory (Sperling's iconic memory, Sternberg scanning, memory
span, memory for binary sequences); numerical cognition (SNARC,
number comparison, number change detection, magnitude estimation);
language and reading (lexical decision, masked priming, the
letter-size illusion, temporal integration in word recognition,
Stroop and picture naming with a voice key, shadowing); geometry and
reasoning (intruder and triangle-completion tasks, mental rotation in
two and three dimensions, the ConceptARC benchmark, Rush~Hour);
psychophysics (contrast detection by QUEST, auditory threshold
estimation); statistical learning (auditory, visual, community
structure); infant looking-time paradigms; motor timing (finger
tapping and tracking, mouse tracking); and neuroimaging localizers
(language localizers in three variants, MEG localizers, standard and
multifocal retinotopic mapping, RSVP of images and of mathematics and
language). Each serves as a functional template, and the collection
as a whole doubles as a corpus from which AI coding assistants
generate new experiments in the same idiom.

All examples and demonstrations are available pre-built for Windows,
macOS and Linux (Section~\ref{single-binary-deployment}), and most
also run in the browser (Section~\ref{browser}). Building them
locally takes three steps: install Go, clone the repository, and run
\texttt{examples/build.sh}.

\section{Comparison with Existing Tools}\label{comparison-with-existing-tools}

Table~\ref{tab:comparison} summarizes key properties of \texttt{goxpyriment}
alongside five other experiment frameworks.

\begin{table*}[tb]
\caption{Comparison of selected experiment frameworks.}
\label{tab:comparison}
\footnotesize
\begin{tabular}{@{}
  >{\raggedright\arraybackslash}p{0.17\textwidth}
  >{\raggedright\arraybackslash}p{0.14\textwidth}
  >{\raggedright\arraybackslash}p{0.14\textwidth}
  >{\raggedright\arraybackslash}p{0.14\textwidth}
  >{\raggedright\arraybackslash}p{0.14\textwidth}
  >{\raggedright\arraybackslash}p{0.14\textwidth}@{}}
\toprule
Property & goxpyriment & Expyriment & PsychoPy & PTB (MATLAB) & jsPsych \\
\midrule
Language        & Go           & Python        & Python            & MATLAB           & JavaScript \\
Deployment      & Single binary & pip/conda    & Standalone or pip & MATLAB license   & CDN / npm \\
GC control      & Explicit disable & No        & No                & N/A              & No \\
VSYNC sync      & Yes (SDL3)   & Yes (SDL2)    & Yes (Pyglet/PTB)  & Yes              & No \\
Hardware triggers & Yes        & Yes           & Yes               & Yes              & No \\
Eye tracking    & EyeLink, Tobii (bridge) & No & ioHub (several vendors) & EyeLink toolbox & WebGazer (webcam) \\
Response device API & Unified  & Device-specific & Device-specific & Device-specific & N/A \\
Event timestamp & SDL3 hardware & pygame timer & iohub/PTB clock  & \texttt{GetSecs} ($\mu$s) & Browser clock \\
Stimulus set    & Moderate     & Moderate      & Extensive         & Extensive        & Moderate \\
GUI builder     & No           & No            & Yes (Builder)     & No               & No \\
Online deployment & Yes (same source, WASM) & No & Partial (PsychoJS) & No           & Yes \\
Reference       & This paper   & \cite{krause2014} & \cite{peirce2019} & \cite{brainard1997} & \cite{deleeuw2023} \\
\bottomrule
\end{tabular}
\end{table*}

\texttt{goxpyriment} resembles Expyriment in its API design: a central
experiment object, a \texttt{present()}-based stimulus API, a
structured trial-loop model, and output to \texttt{.csv}-format files
with metadata. The code for a simple experiment (Parity Decision) is
provided in Listing~\ref{paritycode}. The primary technical differences
are the use of Go rather than Python and the use of SDL3 rather than
the older Pygame/SDL2 stack. From the user's point of view,
\texttt{goxpyriment} adds \emph{stream} objects which, as their name
indicates, allow displaying sequences of stimuli with tightly
controlled timing.

Table~\ref{tab:comparison} necessarily simplifies; each framework has
strengths that a compact matrix cannot convey. PsychoPy, in
particular, offers a graphical Builder interface, a larger and
extensively documented stimulus set, mature online deployment through
PsychoJS, and a large user community. The Psychtoolbox remains a
reference for low-level timing control on well-configured
hardware. Relative to these established tools, \texttt{goxpyriment} is
very young: it has no graphical builder and a more limited stimulus
library, and its browser deployment, while it runs the same source,
carries neither the hardware retrace clock nor the photodiode
validation of the native build. Its distinctive contributions are
therefore deployment simplicity---one binary, or one URL, from one
source---and an explicit timing architecture rather than breadth of
features. The entries in
Table~\ref{tab:comparison} describe design characteristics, not measured
timing.

\subsection{Expandability}

The availability of the source code makes it easy to expand
goxpyriment, e.g., adding support for new hardware. Given a clear
specification for an hardware interface, an human, or an AI coding
assistant can quickly generate Go code for this hardware taking
example from the existing packages.

Goxpyriment is distributed under an Apache-2 license, a free software
license compatible with the GNU GPLv3 according to the Free Software
Foundation (GPLv3 projects can include code licensed under Apache-2).

\section{Limitations and Future Directions}\label{limitations-and-future-directions}

\textbf{Timing is characterised on Linux only.} The timing measurements
made so far were all made under Linux. The
framework builds and runs on macOS and Windows, but real-time
scheduling is requested only on Linux, and no photodiode measurement
has been made on the other two systems; an experimenter who needs
onset guarantees there should run the bundled timing suite on their
own machine before trusting any number.

\textbf{Browser onsets are unvalidated.} The browser build has been
checked for timestamp resolution and frame pacing but not, so far,
against a photodiode; until it is, it should be reserved for paradigms
that tolerate browser-grade onset precision.

\textbf{Eye-tracker calibration.} On the EyeLink, calibration targets
are drawn by the vendor library in its own window rather than by
\texttt{goxpyriment}, so the calibration display cannot yet be styled
like the experiment; the Tobii path already draws its own targets and
the same could be done for the EyeLink. Trackers that publish a
network protocol (GazePoint, Pupil~Labs) could be supported by a
pure-Go client without a bridge.

\textbf{No video recording}. While there is currently some support for
recording audio responses and measuring naming times, the framework
does not currently allow to capture videos (unless one controls a
networked video recorder, which the trigger package supports over
TCP/IP).  As SDL3 supports recording video, this feature might be
implemented in the future.

\textbf{Server-side collection only through JATOS}.  A browser build
can upload its results to a JATOS server (Section~\ref{browser}), but
native builds save their results locally, and there is no built-in
upload to other back-ends, for example an on-line database behind a
REST api.

\textbf{Language barrier.}  Researchers who are unfamiliar with Go,
that is, the vast majority, will need to invest a bit of time in
getting acquainted with the language.  To this end, many web resources
exist, for example, \url{https://go.dev/tour/} and
\url{https://gobyexample.com/}. As Go is a descendant of C, the syntax
will be familiar to people who know a language of the same family,
e.g., Java or Javascript. Finally, the examples and the migration
guide we provide at \url{https://chrplr.github.io/goxpyriment}, are
also meant to ease the transition. AI-Coding tools can help a lot in
porting existing matlab or python code, adapting existing experiments,
or creating new ones.

\clearpage

\section{Declarations}

\subsection{Funding}

No funding was received to assist with this project nor with the preparation of this manuscript.

\subsection{Conflict of interest/Competing interest}

The authors have no competing interests to declare that are relevant to the content of this article.

\subsection{Ethics approval}

Not applicable.

\subsection{Consent to participate}

Not applicable.

\subsection{Consent for publication}

Not applicable.

\subsection{Availability of data and materials}

The raw instrument captures, run logs and analysis scripts of the timing
measurements will be made available with the source repository.

\subsection{Code availability}

The source code of \texttt{goxpyriment} is available under the Apache
License 2.0 at
\url{https://github.com/chrplr/goxpyriment}. All example experiments
described in this article are included in the repository. No
behavioural data from human participants are reported; the empirical
measurements in this article are hardware timing measurements.

\subsection{Authors Contributions}\label{author-contributions}

\textbf{Christophe Pallier}: Conceptualization, Software
implementation, Documentation and Paper Writing, with assistance from
\textbf{Claude Opus 5} and \textbf{Gemini 2.5}. Consistent with
current editorial norms \parencite{natureportfolio2023,
  springernature2023}, Claude and Gemini are listed under Author
Contributions rather than as a named author, because authorship
requires the capacity to take accountability for the work.  Christophe
Pallier takes full responsibility for all content. \textbf{Julie
  Bonnaire} and \textbf{Marie-France Fourcade} set up the timing
measurement system and performed tests.

\clearpage

\printbibliography

@online{golang,
  author       = {{The Go Authors}},
  title        = {Go developpers website},
  year         = {2026},
  url          = {https://go.dev},
  urldate      = {2026-07-26},
  organization = {Google}
}

@online{pygame,
  author       = {{Pygame Developers}},
  title        = {Pygame: Python Game Development Development Library},
  year         = {2026},
  url          = {https://www.pygame.org/},
  urldate      = {2026-07-26},
  organization = {Pygame Community}
}

@online{sdl3,
  author = {{SDL Developers}},
  title = {Simple {D}irect{M}edia {L}ayer, SDL3},
  url = {https://www.libsdl.org/},
  year = {2026},
  urldate = {2026-04-01}
}

@online{gosdl3,
  author = {Bertrand Jung},
  title = {{SDL3} {B}indings for {G}o},
  url = {https://github.com/Zyko0/go-sdl3},
  year = {2026},
  urldate = {2026-04-01}
}

@article{anwylirvine2021,
  author  = {Anwyl-Irvine, Alexander L. and Massonnié, Jessica and Flitton, Adam and Kirkham, Natasha and Evershed, Jo K.},
  title   = {Gorilla in our midst: {An} online behavioral experiment builder},
  journal = {Behavior Research Methods},
  year    = {2021},
  volume  = {52},
  number  = {1},
  pages   = {388--407},
  doi     = {10.3758/s13428-019-01237-x}
}

@article{brainard1997,
  author  = {Brainard, David H.},
  title   = {The {Psychophysics} {Toolbox}},
  journal = {Spatial Vision},
  year    = {1997},
  volume  = {10},
  number  = {4},
  pages   = {433--436},
  doi     = {10.1163/156856897X00357}
}

@article{bridges2020,
  author  = {Bridges, David and Pitiot, Alain and MacAskill, Michael R. and Peirce, Jonathan W.},
  title   = {The timing mega-study: {Comparing} a range of experiment generators, both lab-based and online},
  journal = {PeerJ},
  year    = {2020},
  volume  = {8},
  pages   = {e9414},
  doi     = {10.7717/peerj.9414}
}

@article{deleeuw2015,
  author  = {de Leeuw, Joshua R.},
  title   = {{jsPsych}: {A} {JavaScript} library for creating behavioral experiments in a {Web} browser},
  journal = {Behavior Research Methods},
  year    = {2015},
  volume  = {47},
  number  = {1},
  pages   = {1--12},
  doi     = {10.3758/s13428-014-0458-y}
}

@article{deleeuw2023,
  author  = {de Leeuw, Joshua R. and Gilbert, Rebecca A. and Luchterhandt, Bj{\"o}rn},
  title   = {{jsPsych}: {Enabling} an open-source collaborative ecosystem of behavioral experiments},
  journal = {Journal of Open Source Software},
  year    = {2023},
  volume  = {8},
  number  = {85},
  pages   = {5351},
  doi     = {10.21105/joss.05351}
}

@book{donovan2016,
  author    = {Donovan, Alan A. A. and Kernighan, Brian W.},
  title     = {The {Go} Programming Language},
  publisher = {Addison-Wesley},
  year      = {2016}
}

@article{henninger2022,
  author  = {Henninger, Felix and Shevchenko, Yury and Meijer, Rolf R. and Inzlicht, Michael and Hilbig, Benjamin E.},
  title   = {lab.js: {A} free, open, online study builder},
  journal = {Behavior Research Methods},
  year    = {2022},
  volume  = {54},
  number  = {2},
  pages   = {556--573},
  doi     = {10.3758/s13428-019-01283-5}
}

@article{kleiner2007,
  author  = {Kleiner, Mario and Brainard, David H. and Pelli, Denis},
  title   = {What's new in {Psychtoolbox-3}?},
  journal = {Perception},
  year    = {2007},
  volume  = {36},
  number  = {ECVP Abstract Supplement},
  pages   = {14},
  url = {https://pure.mpg.de/rest/items/item_1790332/component/file_3136265/content}
}

@article{krause2014,
  author  = {Krause, Florian and Lindemann, Oliver},
  title   = {Expyriment: {A} {Python} library for cognitive and neuroscientific experiments},
  journal = {Behavior Research Methods},
  year    = {2014},
  volume  = {46},
  number  = {2},
  pages   = {416--428},
  doi     = {10.3758/s13428-013-0390-6}
}

@misc{natureportfolio2023,
  author       = {{Nature Portfolio}},
  title        = {Tools and technologies: {Artificial} intelligence},
  year         = {2023},
  howpublished = {\url{https://www.nature.com/nature-portfolio/editorial-policies/ai}}
}

@article{peirce2019,
  author  = {Peirce, Jonathan W. and Gray, Jeremy R. and Simpson, Sol and MacAskill, Michael R. and H{\"o}chenberger, Richard and Sogo, Hiroyuki and Kastman, Erik and Lindel{\o}v, Jonas K.},
  title   = {{PsychoPy2}: {Experiments} in behavior made easy},
  journal = {Behavior Research Methods},
  year    = {2019},
  volume  = {51},
  number  = {1},
  pages   = {195--203},
  doi     = {10.3758/s13428-018-01193-y}
}

@misc{springernature2023,
  author       = {{Springer Nature}},
  title        = {Artificial intelligence ({AI}) policy for authors},
  year         = {2023},
  howpublished = {\url{https://www.springernature.com/gp/policies/editorial-policies}}
}

@article{visionegg_2008,
        title = {Vision {Egg}: an open-source library for realtime visual stimulus generation},
        volume = {2},
        issn = {1662-5196},
        shorttitle = {Vision {Egg}},
        url = {https://www.frontiersin.org/journals/neuroinformatics/articles/10.3389/neuro.11.004.2008/full},
        doi = {10.3389/neuro.11.004.2008},
        language = {English},
        urldate = {2026-07-22},
        journal = {Frontiers in Neuroinformatics},
        publisher = {Frontiers},
        author = {Straw, Andrew D.},
        month = nov,
        year = {2008},
}

@article{pyepl_2007,
        title = {{PyEPL}: {A} cross-platform experiment-programming library},
        volume = {39},
        issn = {1554-3528},
        shorttitle = {{PyEPL}},
        url = {https://doi.org/10.3758/BF03192990},
        doi = {10.3758/BF03192990},
        language = {en},
        number = {4},
        urldate = {2026-07-22},
        journal = {Behavior Research Methods},
        author = {Geller, Aaron S. and Schleifer, Ian K. and Sederberg, Per B. and Jacobs, Joshua and Kahana, Michael J.},
        month = nov,
        year = {2007},
        pages = {950--958},
}

@article{levitt1971,
  author  = {Levitt, H.},
  title   = {Transformed up-down methods in psychoacoustics},
  journal = {The Journal of the Acoustical Society of America},
  year    = {1971},
  volume  = {49},
  number  = {2B},
  pages   = {467--477},
  doi     = {10.1121/1.1912375}
}

@article{watson1983,
  author  = {Watson, Andrew B. and Pelli, Denis G.},
  title   = {{QUEST}: A {Bayesian} adaptive psychometric method},
  journal = {Perception \& Psychophysics},
  year    = {1983},
  volume  = {33},
  number  = {2},
  pages   = {113--120},
  doi     = {10.3758/BF03202828}
}

@article{lange2015,
  author  = {Lange, Kristian and K{\"u}hn, Simone and Filevich, Elisa},
  journal = {PLOS ONE},
  year    = {2015},
  volume  = {10},
  number  = {6},
  pages   = {e0130834},
  title = {``{Just} {Another} {Tool} for {Online} {Studies}'' ({JATOS}): {An} {Easy} {Solution} for {Setup} and {Management} of {Web} {Servers} {Supporting} {Online} {Studies}},
        issn = {1932-6203},
        shorttitle = {``{Just} {Another} {Tool} for {Online} {Studies}'' ({JATOS})},
        url = {https://journals.plos.org/plosone/article?id=10.1371/journal.pone.0130834},
        doi = {10.1371/journal.pone.0130834}
}

\clearpage
\appendix

\section{Supplementary Material: Timing Validation in Detail}\label{app:timing}

This appendix documents the apparatus and the commands of the timing
measurements. Their results are in progress.

\subsection{Apparatus}

\paragraph{Machines.} Three, chosen to span the range of hardware a laboratory
is likely to own rather than to represent it: a single-board computer, a
workstation with a professional GPU, and a current laptop. All ran Linux with
SDL~3.4.0, with the garbage collector suspended for the duration of each
measurement loop and the process at \texttt{SCHED\_FIFO} priority~50.

\begin{center}
\scriptsize
\begin{tabular}{llll}
\hline \hline
Host & Machine & Graphics & Audio \\
\hline
\texttt{omv-pi4}   & Raspberry Pi 4B          & V3D 4.2.14.0                & on-board codec \\
\texttt{pcstim7}   & workstation              & AMD Radeon Pro W5700 (radeonsi, navi10) & Focusrite Scarlett Solo 3rd Gen \\
\texttt{is158520}  & Dell Precision 5490      & Intel Arc Graphics (Meteor Lake) & on-board (PipeWire) \\
\hline
\end{tabular}
\end{center}

\paragraph{Displays.} A single Dell 1905FP (1280$\times$1024, nominal 60.0197~Hz)
was shared between \texttt{pcstim7} and \texttt{is158520}, connected by a
DisplayPort-to-HDMI adapter on the first and a USB-C-to-HDMI hub on the second.
The 5490's built-in panel (2560$\times$1600) was measured separately, and the
Pi drove an HDMI display. Using the same physical monitor on two machines is
what allows the panel's contribution to be separated from the software's.

\paragraph{Instruments.} A Black Box Toolkit, and a Digilent Analog Discovery~3
sampling two channels at 100~kS/s on one timebase. Channel~1 carried a
photodiode taped to the screen over the stimulus square; channel~2 carried the
trigger line, or the audio line output for the audio-visual runs. Because both
channels share one acquisition and one clock, the instrument's timebase cancels
in any interval measured between them.

Two properties of the AD3 rig cost us time and are worth stating. Its analogue
inputs are differential, so a lead that leaves the negative input floating reads
mains hum rather than signal; and the device must be on a root USB port---behind
a laptop dock it enumerates normally, accepts every configuration, and then
delivers about 20 samples per second before stalling.

\begin{figure}[htb]
\centering
\includegraphics[width=0.58\linewidth]{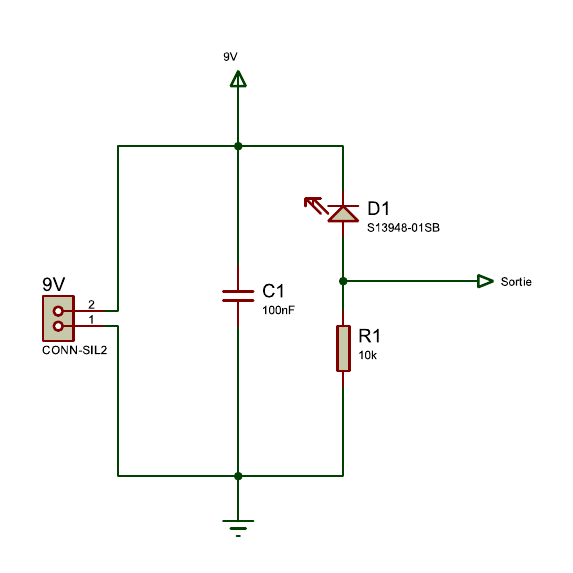}
\caption{The photodiode detector used on channel~1. D1 is reverse-biased from a
  9~V battery and its photocurrent develops across the 10~k$\Omega$ load R1, so
  the output at \emph{Sortie} rises with screen luminance; C1 decouples the
  supply. Running the detector from a battery keeps it off the mains ground of
  the machine under test.}
\label{fig:photodiode}
\end{figure}

\paragraph{Photodiode.} Figure~\ref{fig:photodiode} gives the detector circuit.
It is a passive design on purpose: with no amplifier in the path there is no
amplifier settling time to disentangle from the panel's. The load resistance and
the junction-plus-cable capacitance, tens of picofarads, put the detector's own
time constant well under a microsecond---below the 10~$\mu$s sample interval,
and three orders of magnitude faster than the millisecond panel transitions it
is used to measure.

Onset and offset criteria are applied per trial against that trial's own dark
and bright levels rather than against a fixed voltage, so a slow drift in
backlight output does not leak into the transition times.

\paragraph{Triggers.} A DLP-IO8 USB-serial digital I/O box on \texttt{pcstim7} and
\texttt{is158520}, and a directly-driven GPIO line on the Pi.

\subsection{Runs and commands}

Each run presents a square for 12 frames in a 30-frame (500~ms) cycle, emitting
a TTL pulse on the flip that turns it on. The commands below are those recorded
in each run's own \texttt{-info.txt}; the AD3 capture is started first, from a
second machine or a second terminal, and outlives the run.

\begin{lstlisting}[basicstyle=\ttfamily\scriptsize,numbers=none,breaklines=true]
# 1. start the capture (root USB port; --save-raw is needed for envelope work)
ad3-capture --seconds 545 --rate 1e5 --channels 1,2 --save-raw --out run.npz

# 2. run the measurement (examples as actually invoked)
Timing-Tests -test display -duration-s 300
Timing-Tests -test av -cycles 1010 -no-sound -trigger-ms 15 \
             -trigger-device dlpio8 -port /dev/serial/by-id/usb-DLP_Design_DLP-IO8_...
Timing-Tests -test av -cycles 1010 -no-sound -trigger-ms 15 \
             -trigger-device dlpio8 -d 1 -port /dev/serial/by-id/...   # external display
Timing-Tests -test av -cycles 500 -audio-frames 1024 -trigger-device dlpio8 -trigger-ms 15

# 3. extract events, and fit the host-to-instrument clock relation
./cmd/ad3-to-events.py run.npz events.csv       # also prints the panel rise/fall
./_build/timing-drift events.csv Timing-Tests_sub-000_date-....csv

# 4. plot the timing figures
./cmd/plot-timing-figures.py <capture-dir> paper/figures
\end{lstlisting}

\begin{figure}[htb]
\begin{lstlisting}[basicstyle=\ttfamily\tiny,numbers=none,frame=single,
                 breaklines=true,columns=fullflexible]
# --EXPERIMENT INFO
# e mainfile: Timing-Tests
# e cmdline: Timing-Tests -test display -duration-s 3 -w
# e start_time: 20260818-142047
# e hostname: is158520
#   [...]
# e os: linux/amd64
# e framework: goxpyriment (devel) --- see http://chrplr.github.io/goxpyriment
#   [...]
# --SUBJECT INFO
# s id: 0
#   [...]
# --SYSTEM INFO
# sys sdl_version: 3.4.0
# sys video_driver: wayland
# sys renderer: opengl
# sys gl_renderer: Mesa Intel(R) Arc(tm) Graphics (MTL)
#   [...]
# sys fullscreen: false
# sys vsync: 1
# sys vblank_backend: not requested (set GOXPY_VBLANK=on to anchor onsets on kernel vblanks)
# sys refresh_nominal_hz: 60.0190
# sys refresh_measured_hz: 60.0025
# sys audio_driver: pipewire
# sys audio_device: Meteor Lake-P HD Audio Controller Speaker
#   [...]
# sys audio_buffer_frames: 1024
# sys sched_policy: SCHED_FIFO
# sys sched_realtime: true
# sys sched_priority: 50
#   [...]
# --HOST INFO
# host machine: Dell Inc. Precision 5490
# host machine_type: laptop
# host os: Ubuntu 26.04 LTS
# host kernel: 7.0.0-29-generic x86_64
# host desktop: ubuntu:GNOME
# host cpu: Intel(R) Core(TM) Ultra 7 165H
# host cpu_topology: 16 cores / 22 threads
# host cpu_mhz: 1667 (max 4700)
# host ram: 30.04 GiB (40% used at start)
# host gpu0: Intel Corporation Meteor Lake-P [Intel Arc Graphics] [driver i915]
# host gpu1: NVIDIA Corporation AD107GLM [RTX 2000 Ada Generation Laptop GPU] [driver nvidia]
# host audio_card: sof-soundwire [driver snd_soc_sof_sdw]
# host audio_server: PipeWire 1.6.2
# host alsa: k7.0.0-29-generic
# --DISPLAY INFO
# d id: 3
# d name: Dell Inc. 19"
# d native_resolution: 1280x1024
#   [...]
# d refresh_rate_hz: 60.0200
#   [...]
# d bounds: 0,0 1280x1024
# gc=suspended
#   [...]
# test=display duration-s=3.0 warmup=10
#   [...]
# sys pacing: presents=171 blocked=171 early=93 paced=0 vblank_held=0 (0.0 % paced)
#              wait_mean=0.000 ms wait_max=0.000 ms frame=16.661 ms hold_frac=0.000 class=blocking
# e end_time: 20260818-142052.248
# e duration: 00:00:04.509
\end{lstlisting}
\caption{The companion \texttt{-info.txt} written beside every session's CSV,
  from a short run on the Dell Precision 5490 (\texttt{is158520}; abridged; each
  \texttt{[...]} marks omitted lines of the same kind, nothing is reworded).
  \texttt{sys} lines are what SDL reports about what it opened; \texttt{host}
  lines are what the operating system says about the machine. The CSV beside it
  carries no comment lines at all, so it loads into a spreadsheet or statistics
  package unmodified. Every field here is written without the experimenter doing
  anything, which is the point: the compositor, the sound server version and the
  scheduling policy obtained are all conditions that can affect timing, and none
  of them can be recovered from the data afterwards.}
\label{fig:infofile}
\end{figure}

To change the display stack, the desktop session was left running for the
Wayland and X11 conditions, and stopped in favour of a virtual console for the
KMS/DRM conditions; nothing else was altered between the members of a pair.

\subsection{Results}

in progress...

\subsection{Availability}

The measurement suite is \texttt{tests/Timing-Tests/} in the repository, and the
analysis tools used here are \texttt{cmd/ad3-to-events.py},
\texttt{cmd/extract-av-sync.py}, \texttt{cmd/plot-durations.py} and
\texttt{cmd/plot-timing-figures.py}. Capture is performed with
\texttt{ad3-capture} (\url{https://github.com/chrplr/ad3-capture}), which is
released separately because it is useful independently of this framework. Every
run writes its configuration---display mode, audio device and buffer, scheduling
policy, and the counts of each pacing branch---into a companion
\texttt{-info.txt}, and the source of each cycle's onset timestamp into its CSV.

\end{document}